\documentclass[11pt, a4paper]{article}
\usepackage{}

\usepackage{indentfirst}
\usepackage{lineno}
\usepackage{graphicx}
\usepackage{ae}
\usepackage{amsmath}
\usepackage{amssymb}
\usepackage{latexsym}
\usepackage{url}
\usepackage{epsfig}
\usepackage{cite}
\usepackage{mathrsfs}
\usepackage{amsfonts}
\usepackage{amsthm}
\usepackage{float}
\usepackage{booktabs}
\usepackage{subfigure}
\usepackage{multirow}

\usepackage{color}

\long\def\delete#1{}

\newcommand{\be}{\begin{equation}}
\newcommand{\ee}{\end{equation}}
\newcommand{\ben}{\begin{equation*}}
\newcommand{\een}{\end{equation*}}
\newcommand{\bea}{\begin{eqnarray}}
\newcommand{\eea}{\end{eqnarray}}
\newcommand{\bean}{\begin{eqnarray*}}
\newcommand{\eean}{\end{eqnarray*}}

\numberwithin{equation}{section}

\title{A mixed clustering coefficient centrality for identifying essential proteins\thanks{Supported by the National Natural Science Foundation of China (No.11361033) and the Natural Science Foundation of Gansu Province (No.1212RJZA029).}}

\author{Pengli Lu\thanks{Corresponding author.   E-mail
addresses: lupengli88@163.com (\textbf{P. Lu}), yujingjuanmercy@163.com (\textbf{J. Yu}).} \;and\; JingJuan Yu
\\
\footnotesize{School of Computer and Communication, Lanzhou University of Technology, Lanzhou, 730050, Gansu, P.R. China}}

\date{}

\begin{document}

\openup 0.5\jot
\maketitle

\begin{abstract}
Essential protein plays a crucial role in the process of cell life. The identification of essential proteins can not only promote the development of drug target technology, but also contribute to the mechanism of biological evolution. There are plenty of scholars who pay attention to discovering essential proteins according to the topological structure of protein network and biological information. The accuracy of protein recognition still demands to be improved. In this paper, we propose a method which integrate the clustering coefficient in protein complexes and topological properties to determine the essentiality of proteins. First, we give the definition of In-clustering coefficient ($IC$) to describe the properties of protein complexes. Then we propose a new method, complex edge and node clustering coefficient ($CENC$) to identify essential proteins. Different Protein-Protein Interaction (PPI) networks of Saccharomyces cerevisiae, MIPS and DIP are used as experimental materials. Through some experiments of logistic regression model, the results show that the method of $CENC$ can promote the ability of recognizing essential proteins, by comparing with the existing methods $DC$, $BC$, $EC$, $SC$, $LAC$, $NC$ and the recent method $UC$.
\bigskip

\bigskip

\end{abstract}

\section{Introduction}
Protein is a crucial component of all cells and organizations. It is considered as essential proteins that the proteins necessary to maintain the life of the organism. Not only can essential proteins promote the development of drug target technology, but also help the study of biological evolution mechanism \cite{1}. Removing the essential proteins can lead to cell death or inability to replicate and reproduce \cite{2}. The recognition and protection of essential proteins are the basis of drug development, which provide valuable theories and methods for the diagnosis of diseases, drug design, etc. \cite{3}.

In biology, the identification methods of essential proteins mainly rely on biological experiments, such as conditional knockouts \cite{4}, RNA interference \cite{5}, and single gene knockouts \cite{6}, coupled with the survival ability of infected organisms being tested. These biological experimental results are clear and effective, but consume amounts of time, costs and resources. With the improvement of prior technology, several protein protein interaction(PPI) networks are generated \cite{43,44}. Nowadays, it has been a crucial research direction in the field of bioinformatics for predicting essential proteins from a large number of biological experiments by using the theory of technology from PPI networks. The methods for identifying essential proteins can be divided into several categories.

Based on the centrality-lethality rule which put forward by Jeong H M et al., the essentiality of proteins is associated with the topological structure in PPI networks \cite{7}. Thus, a large number of scholars have proposed many indicators based on topological centrality \cite{10,47}. Some of them considered the topological of nodes in networks, such as degree centrality ($DC$) which considers the connection nodes \cite{10,14,15}, betweenness centrality ($BC$) which considers the global characteristic \cite{11,38}, subgraph centrality ($SC$) \cite{16}, local average centrality ($LAC$) \cite{17}, eigenvector centrality ($EC$) \cite{19}, information centrality ($IC$) \cite{20} and closeness centrality ($CC$) \cite{13}, and topology potential-based method ($TP$) \cite{34}. Some of them considered the topological of edges in networks, including edge clustering coefficient ($ECC$) \cite{35}, improved node and edge clustering coefficient ($INEC$) \cite{36}, integrated edge weights ($IEC$)\cite{45} and network centrality ($NC$) \cite{46}. CytoNCA, an app of Cytoscape for analyzing the centrality methods, have been a valuable tool to identify the essentiality of proteins \cite{42}.

With the increase of high-throughput biological data, scholars have tried to combine with biology information to improve the accuracy of identifying essential proteins. Considering the functional annotations of genes, a weighted protein-protein network is constructed, by combining edge clustering coefficients with gene expression data correlation coefficients, a method of $PeC$ is proposed \cite{24}. There is an $esPOS$ method that uses the information of gene expression and subcellular localization \cite{21}. $SPP$ method is based on sub-network division and sequencing by integrating subcellular positioning \cite{12}. Extended pareto optimality consensus model ($EPOC$) mixes neighborhood closeness centrality and Orthology information together \cite{39}. Go terms information is also used to predict essential proteins such as $RSG$ method \cite{25}.

There are some studies who recognize essential proteins from the perspective of complexes and functional modules. Hart G T et al. find that the essentiality is an attribute of the protein complex and the protein complexes often determine the essential proteins \cite{22}. Li et al. prove that the frequency of the essential protein that occurs in the complex is higher than in the whole network \cite{21,41}. Luo J W et al. propose a method of ($LIDC$), combining the local interaction density and protein complexes for predicting essential proteins \cite{37}. Li et al. propose united complex centrality ($UC$) which takes into account the frequency of protein appeared in the complex and edge properties \cite{23}.

In this paper, considering the protein complexes information and topological properties, a new method of complex edge and node clustering coefficient ($CENC$) is proposed to identify essential proteins. To assess the quality of $CENC$ method, different datasets of Saccharomyes cerevisiae, MIPS and DIP are applied. By the comparison of seven existing methods, containing $DC$, $BC$, $EC$, $SC$, $LAC$, $NC$ and $UC$, experiment results show that our method can be more effective in determining the essentiality of proteins than existing measures.


\section{New Centrality: $CENC$}\label{Se2}
An undirected simple graph $G(V, E)$ can be used to express a network of protein interaction. Proteins can be regarded as nodes set $V$ of a network and the connections between two proteins can be regarded as edges set $E$. In this study, we present a new method of complex edge and node clustering coefficient $CENC$ to judge the essentiality of proteins by combining the feature of protein complex and topology of nodes and edges. The basic considerations of $CENC$ are as follows: (1) The essential proteins appear in complexes can be more frequency. (2) Both the topology of node and edge are important factors to affect the essentiality of proteins.

First, we present a classical method of clustering coefficient ($C$) \cite{48}.
    \begin{equation}\label{adjmatix}
    \begin{aligned}
    \begin{split}
    C(v)=\frac{2E_{v}}{k_{v}(k_{v}-1)}
    \end{split}
    \end{aligned}
    \end{equation}
where $E_{v}$ is the actual number of edges shared with local neighbors of node $v$, $k_{v}$ is the degree of node $v$.

Then, a clustering coefficient of a node to an edge was generalized by Radicchi et al. \cite{49}. The edge clustering coefficient ($ECC$) is defined as \cite{35}.
    \begin{equation}\label{adjmatix}
    \begin{aligned}
    \begin{split}
    ECC_{v,u}=\frac{z_{v,u}}{min(k_{v}-1,k_{u}-1)}
    \end{split}
    \end{aligned}
    \end{equation}
where $z_{v,u}$ is the number of triangles that includes the edge $e(v,u)$ in network. $k_{v}$ and $k_{u}$ are the degrees of node $u$ and node $v$, respectively.

Based on the numbers of connection edges for a node and the clustering coefficient of each edge, the sum of edge clustering coefficients $NC$ is proposed \cite{46}.
    \begin{equation}\label{adjmatix}
    \begin{aligned}
    \begin{split}
    NC(v)=\sum_{u\in N_{v}}ECC(v,u)
    \end{split}
    \end{aligned}
    \end{equation}
where $N_{v}$ denotes the set of all neighbors of node $v$.

Further more, we propose a new definition In-clustering coefficient ($IC$) which combine the feature in complexes.
    \begin{equation}\label{adjmatix}
    \begin{aligned}
    \begin{split}
    IC(v)=\sum_{i\in ComplexSet(v)}C(v)_{i}
    \end{split}
    \end{aligned}
    \end{equation}
A subset of protein complexes that containing protein $v$ can be represented as $ComplexSet(v)$, the value of $C(v)$ for the $i_{th}$ protein complex which belongs to $ComplexSet(v)$ can be represented as $C(v)_{i}$.

Now, based on the definition that we described above, we propose our new method complex edge and node clustering coefficient ($CENC$) for estimating the essentiality of proteins.
    \begin{equation}\label{adjmatix}
    \begin{aligned}
    \begin{split}
    CENC(v)=a*\frac{IC(v)}{IC_{max}}+b*\frac{NC(v)}{NC_{max}}+c*\frac{C(v)}{C_{max}}
    \end{split}
    \end{aligned}
    \end{equation}
where $a$, $b$, $c$ are random factors ranging from $1$ to $10$. Under the amounts of experiments, we can get the best result of the method $CENC$ when $a$, $b$ and $c$ are 10, 1 and 1, respectively.


\section{Experimental data and assessment methods}\label{Se3}
\subsection{Experimental data}

The experiment data are conducted from Saccharomyes cerevisiae, whose proteins are more complete. Two sets of PPI network data MIPS \cite{27} and DIP \cite{26} are used. In the protein network, all self-interactions and repetitive interactions are deleted as a data preprocessing of these PPIs. Specific properties for these two networks are presented in the Table 1. The MIPS network includes 4546 proteins and 12319 interactions, whose clustering coefficient is about 0.0879. In the DIP network, there are 5093 proteins and 24743 interactions, whose clustering coefficient is about 0.0973. The known essential proteins are derived from four databases: MIPS \cite{40}, SGD (Saccharomyces Genome Database) \cite{33}, SGDP (Saccharomyces Genome Deletion Project) \cite{4}, and DEG (Database of Essential Genes) \cite{27}. The protein complex set is from CM270 \cite{40}, CM425 \cite{29}, CYC408 and CYC428 datasets \cite{30,31} which can gain from \cite{21}, containing 745 protein complexes (including 2167 proteins).

\begin{table}[!htbp]
\centering
\caption{Data details of the two protein networks: DIP, MIPS}
\begin{tabular}{|c|c|c|c|c|c|c|}
\hline
Dataset& Proteins& Interactions& Average degree& Essential proteins& Clustering coefficient\\\hline
MIPS & 4546 & 12319 & 5.42 & 1016 & 0.0879  \\\hline
DIP & 5093 & 24743 & 9.72 & 1167 & 0.0973  \\\hline
\end{tabular}
\end{table}

\subsection{Assessment methods}

According to the values of $CENC$, proteins are sorted in descending orders. First, some numbers of top proteins in sequence are selected as predictive essential proteins, then compare them with the real essential proteins. This allows us to know the quantity of true essential proteins. Therefore, the sensitivity ($SN$), specificity ($SP$), F-measure ($F$), and accuracy ($ACC$), positive predictive value ($PPV$), negative predictive value ($NPV$) can be calculated [28,29].

The following are the formulas for calculating these six statistical indicators.

Sensitivity: $$SN=\frac{TP}{TP+FN}$$

Specificity:
$$SP=\frac{TN}{TN+FP}$$

Positive predictive value:
$$PPV=\frac{TP}{TP+FP}$$

Negative predictive value:
$$NPV=\frac{TN}{TN+FN}$$

F-measure:
$$F=\frac{2*SN*PPV}{SN+PPV}$$

Accuracy:
$$ACC=\frac{TP+TN}{P+N}$$
where $TP$ stands for the quantity of true essential proteins which are correctly selected as essential proteins. $FP$ is the quantity of nonessential proteins which are incorrectly selected as essential. $TN$ is the quantity of nonessential proteins which are correctly selected as nonessential. $FN$ is the quantity of essential proteins which are incorrectly selected as nonessential. $P$ and $N$ stand for the sum number of essential and nonessential proteins, respectively.


\section{Results}\label{Se4}

\subsection{Comparison with other centrality measures}

\begin{figure}[htbp]
\centering
\includegraphics[width=5in]{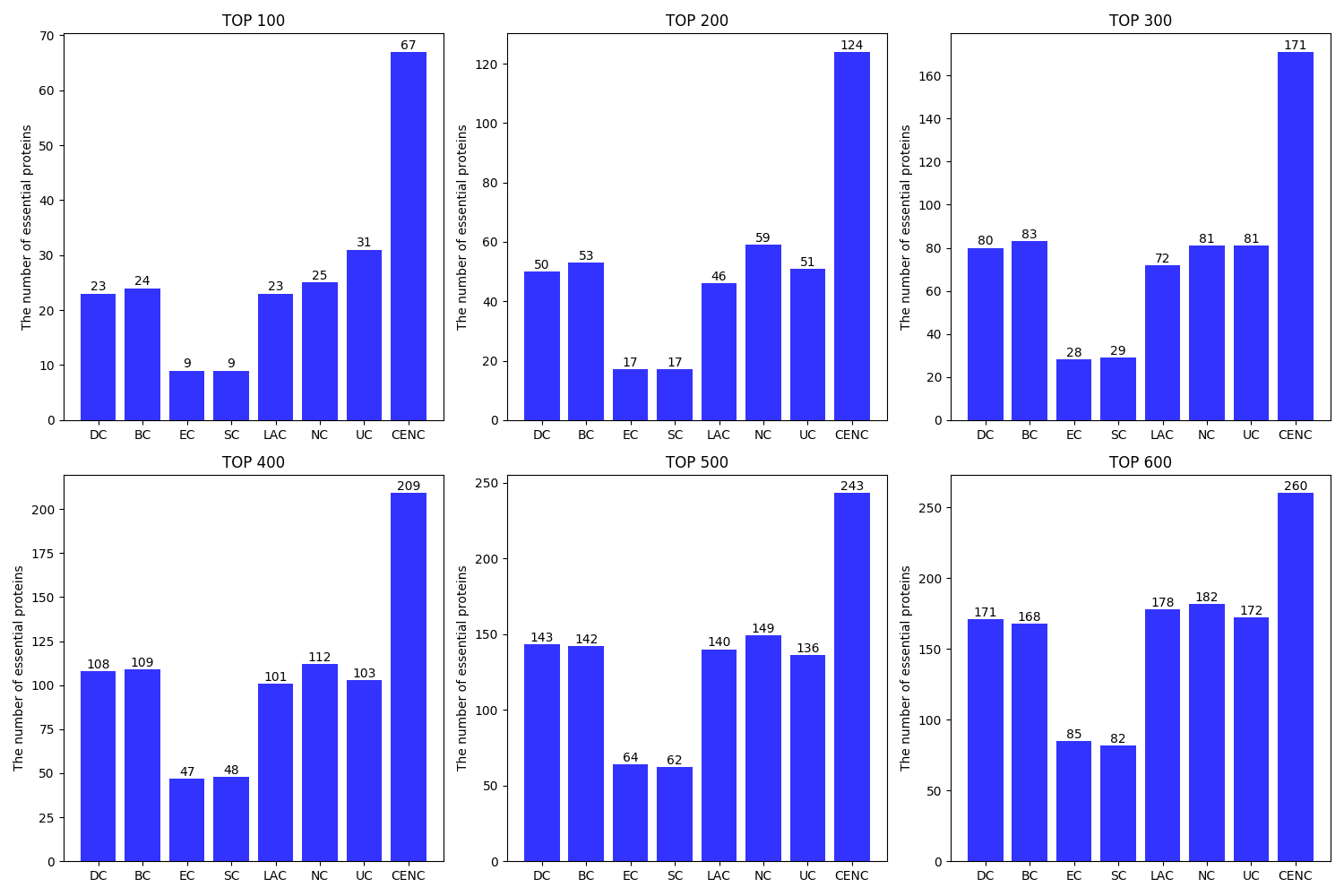}
\caption{The quantity of true essential proteins determined by $CENC$ and other seven previously methods from the MIPS network.}
\centering
\end{figure}

\begin{figure}[htbp]
\centering
\includegraphics[width=5in]{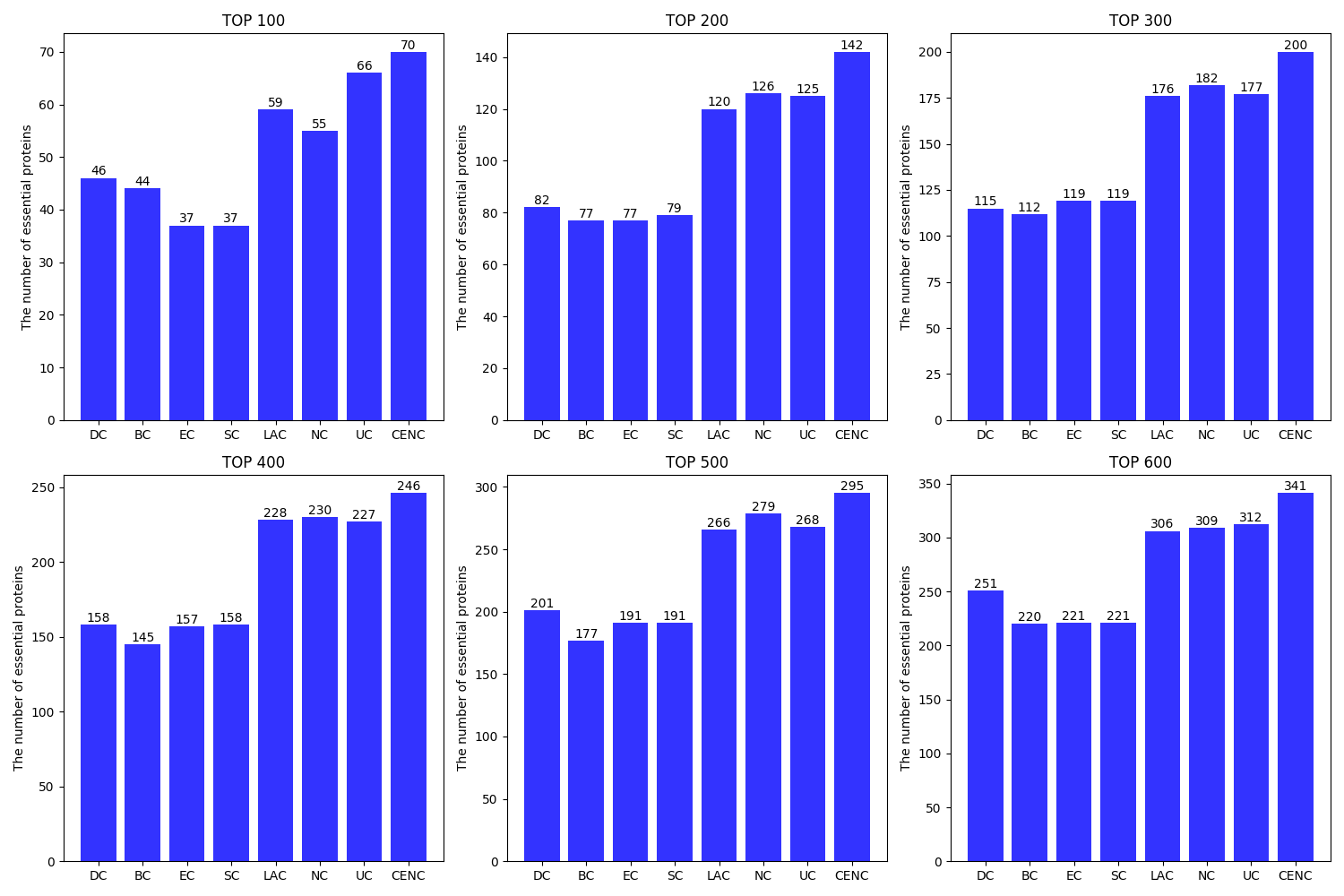}
\caption{The quantity of true essential proteins determined by $CENC$ and other seven previously methods from the DIP network.}
\centering
\end{figure}


We follow the principle of ``sorting-screening" to evaluate the performance of $CENC$. Comparisons of $CENC$ method with other seven previous measures: degree centrality ($DC$) \cite{10}, betweenness centrality ($BC$) \cite{11,38}, eigenvector centrality ($EC$) \cite{19}, subgraph centrality ($SC$) \cite{16}, local average centrality ($LAC$) \cite{17}, network centrality ($NC$) \cite{46}, united complex centrality ($UC$) \cite{23} are carried out in the MIPS and DIP datasets. To be specific, proteins are sorted in descending order on the basis of their values of $CENC$ and other seven previous measures. Then, predictive essential proteins are chosen according to the top 100, 200, 300, 400, 500, and 600 proteins. Finally, by comparing with the known essential proteins, the quantity of true essential proteins among these predictive essential proteins can be obtained. The experimental results of these measures are shown in Figs. 1-2.

From Fig. 1, the quantity of true essential proteins judged by $CENC$ are 67, 124, 171, 209, 243 and 260 from the top 100 to the top 600, respectively, being the best among the eight methods in MIPS network.
Although the method of $UC$ has good performance in the yeast PPI network, it is still poor in MIPS network. Among these seven proposed methods, $SC$ is the lowest indicator of recognition essential proteins. Compared to the method of $SC$, our method $CENC$ improves the rate of 86.56\%, 86.29\%, 83.04\%, 77.03\%, 74.49\%, 68.46\% in the top 100 to top 600, respectively. When choose the best performance for each top, the $CENC$ method can still obtain 53.73\%, 52.42\%, 51.46\%, 46.41\%, 38.68\% and 30\% improvements in predicting essential proteins.

From Fig. 2, it can be seen that the $CENC$ method performs better than existing methods of $DC$, $BC$, $EC$, $SC$, $LAC$, $NC$ and $UC$ in DIP network. Compared with the best result among these seven methods, the true essential proteins determined by $CENC$ method are increased by 4, 16, 18, 16, 16 and 29 from the top 100 to the top 600, respectively. Moreover, the quantity of essential proteins are much more than previous method including $DC$, $BC$, $SC$ and $EC$.


\begin{table}[htbp]
\centering
\small
\caption{\small Comparison the results of sensitivity($SN$), specificity($SP$), F-measure($F$), positive predictive value($PPV$), negative predictive value($NPV$), and accuracy($ACC$) of $CENC$ and other seven previous algorithms.}
\normalsize
\begin{tabular}{|c|c|c|c|c|c|c|c|}
\hline
\multicolumn{0}{|c|} {Dataset}& Methods & SN & SP & PPV & NPV & F-measure & ACC \\\hline
 \cline{1-8}
\multirow{8}*{MIPS} & DC & 0.254&0.803& 0.291 &0.772 &0.271 &0.671\\
\cline{2-8}         & BC &0.197&0.796 &0.278&0.716&0.231 &0.629\\
\cline{2-8}         &EC &0.139&0.773&0.163&0.738&0.150&0.620\\
\cline{2-8}         &SC &0.138 & 0.773& 0.162 & 0.739 &0.149& 0.620\\
\cline{2-8}         &LAC &0.271 & 0.812& 0.314& 0.779& 0.291& 0.682\\
\cline{2-8}         &NC &0.281&0.814 & 0.325&$ 0.781$ & $ 0.302$&0.686\\
\cline{2-8}         &UC&0.271 & 0.812 & 0.314 &0.778 & 0.291& 0.682\\
\cline{2-8}         &CENC &$\mathbf{0.317}$&$\mathbf{0.827}$&$\mathbf{0.368}$&$\mathbf{0.792}$&$\mathbf{0.341}$&$\mathbf{0.704}$\\
\cline{1-8}
\multirow{8}*{DIP}  & DC & 0.353 & 0.834 & 0.409 & 0.80& 0.379 & 0.716\\
\cline{2-8}         & BC & 0.308 & 0.823 & 0.361 & 0.785& 0.333 &0.70\\
\cline{2-8}         &EC & 0.323& 0.824& 0.374 & 0.789 & 0.347 &0.701\\
\cline{2-8}         &SC & 0.316 & 0.822 & 0.366& 0.787 & 0.339 &0.698\\
\cline{2-8}         &LAC &0.405 & 0.852 & 0.472 & 0.815 & 0.436 & 0.743\\
\cline{2-8}         &NC &0.40 &0.850 & 0.463& 0.813&0.428 &0.739\\
\cline{2-8}         &UC &0.391 &0.850 & 0.458& 0.811&0.422 &0.737\\
\cline{2-8}         &CENC &$\mathbf{0.422}$&$\mathbf{0.858} $&$\mathbf{0.491}$ &$\mathbf{0.820}$&$\mathbf{0.454}$ & $\mathbf{0.751}$\\
\hline
\end{tabular}
\end{table}

\begin{figure}[htbp]
\centering
\includegraphics[width=4.5in]{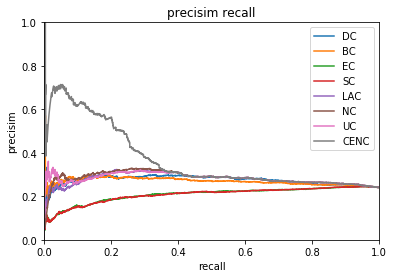}
\caption{Precision and recall curves of $CENC$ and other seven methods for MIPS network.}
\centering
\end{figure}

\begin{figure}[htbp]
\centering
\includegraphics[width=4.5in]{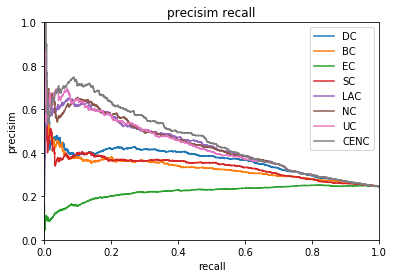}
\caption{Precision and recall curves of $CENC$ and other seven methods for DIP network.}
\centering
\end{figure}
\subsection{Evaluation of six statistical methods and the precision-recall curves}

The six statistical methods are used to evaluate the indicator of $CENC$ as well as other seven identification measures, mentioned in the Assessment methods Section. Proteins are sorted from high to low order on the basis of their values of these methods; Then, the top proteins of 20 percent are taken into account as predictive essential proteins, the remaining 80 percent can be considered as candidates for nonessential proteins. On the two different networks, the comparisons among the values of $CENC$ and other seven measures are executed, showing in Table 2. For DIP network, these six statistic values for $CENC$ are higher than other previous measures, which show that $CENC$ has a better prediction accuracy. For MIPS network, these values of $SN$, $SP$, $PPV$, $NPV$, $F-measure$ and $ACC$ determined by $CENC$ are 0.317, 0.827, 0.368, 0.792, 0.341 and 0.704, respectively, being higher than previous proposed methods $DC$, $BC$, $EC$, $SC$, $LAC$, $NC$ and $UC$. These results indicate that $CENC$ method has a better performance than the existing seven methods.

In addition, the Precision-Recall curve, a statistical method for evaluating stability, can be used for $CENC$ method and other previous seven measures which defined as follows:
$$Precision(n)=\frac{TP(n)}{TP(n)+FP(n)}$$
$$Recall(n)=\frac{TP(n)}{TP(n)+FN(n)}$$
where the definitions of $TP$, $FP$, $FN$ are depicted in the Assessment method Section. The results are revealed in Fig. 3 and Fig. 4. In DIP network, our method of $CENC$ has a better performance than the other methods. The same results are shown in MIPS network.

\subsection{Validation by the Receiver Operating Characteristic ($ROC$) curve and $AUC$}
The Receiver Operating Characteristic ($ROC$) is a valuable tool to measure the imbalance in classification \cite{50}. It is used to evaluate the pros and cons of a binary classifier. Predicting essential proteins can be regard as a two-classification case. Their definitions are as follows:
$$TPR(n)=\frac{TP(n)}{TP(n)+FN(n)}$$
$$FPR(n)=\frac{FP(n)}{FP(n)+TN(n)}$$
where the meanings of $TP$, $FP$, $FN$ and $TN$ are described in the Assessment method Section. As shown in Figs. 5-6, The $ROC$ curve of $CENC$ is slightly higher than that of the other seven methods, indicating the method of $CENC$ is more effective.

To further reveal the experimental results of the $ROC$ curves, the area under the $ROC$ curves is used to quantitatively analysis the results, generally called $AUC$. The $AUC$ results are shown in Table 3. The values of $CENC$ method are much more than the previous existing methods.

\begin{figure}[htbp]
\centering
\includegraphics[width=5in]{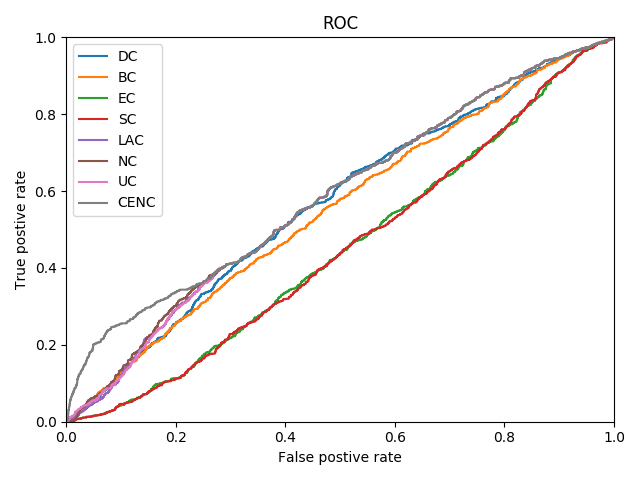}
\caption{$ROC$ curves of $CENC$ and the other seven methods for the MIPS network.}
\centering
\end{figure}
\begin{figure}[htbp]
\centering
\includegraphics[width=5in]{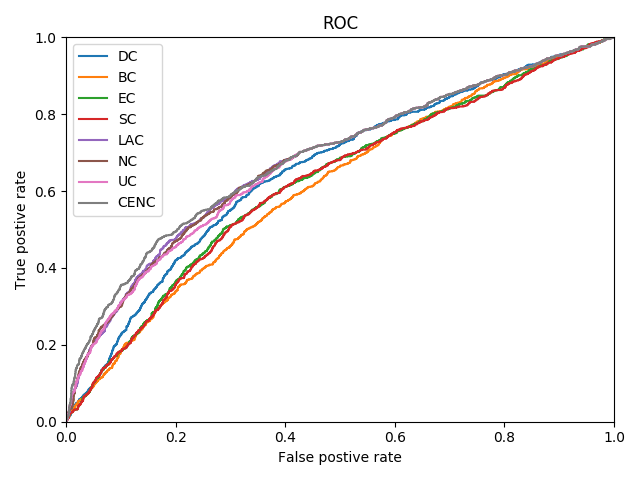}
\caption{$ROC$ curves of $CENC$ and the other seven methods for the DIP network.}
\centering
\end{figure}


\begin{table}[!htbp]
\centering
\caption{\footnotesize $AUC$ values of $CENC$ and other seven methods in  MIPS and DIP networks}
\vspace{0.3em}
\begin{tabular}{|c|c|c|c|c|c|c|c|c|}
\hline
Methods & DC& BC& EC& SC& LAC & NC & UC & CENC \\\hline
MIPS & 0.289 & 0.277 & 0.225 & 0.277 & 0.289 & 0.289 &0.289&0.300\\\hline
DIP & 0.327 & 0.307 & 0.312 & 0.340 & 0.340 & 0.339 & 0.331 & 0.340 \\\hline
\end{tabular}
\end{table}

\subsection{Evaluation of jackknife methodology}
The jackknife methodology was developed by Holman et al., being an effective universal prediction method [30]. The X-axis represents the quantity of selected predictive essential proteins after sequencing, and the Y-axis represents the quantity of true essential proteins in the selected proteins. First, according to the predicted value, proteins are sorted in descending order. Then we choose predictive essential proteins from top 0 to top 800 in each dataset. Last, the jackknife curve is drawn based on the accumulated quantity of real essential proteins. From Fig. 7 and Fig. 8, we can see that the prediction efficiency for $CENC$ method is higher than that of other seven centrality measures on the MIPS and DIP networks. Consequently, the jackknife curves reveal that our method $CENC$ is an effective approach for predicting essential proteins.

\begin{figure}[htbp]
\centering
\includegraphics[width=6.5in]{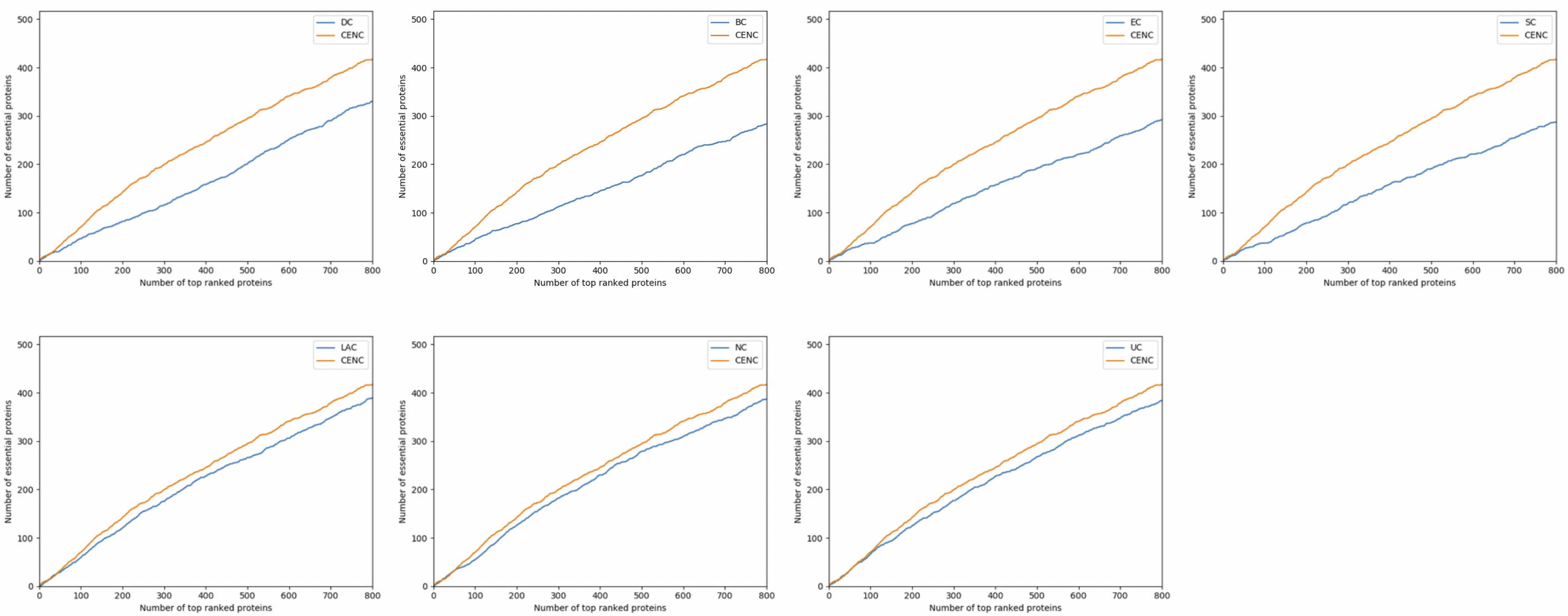}
\caption{The performances of $CENC$ and other seven centrality measures on the DIP network are evaluated by a jackknife methodology.}
\centering
\end{figure}

\begin{figure}[htbp]
\centering
\includegraphics[width=6.5in]{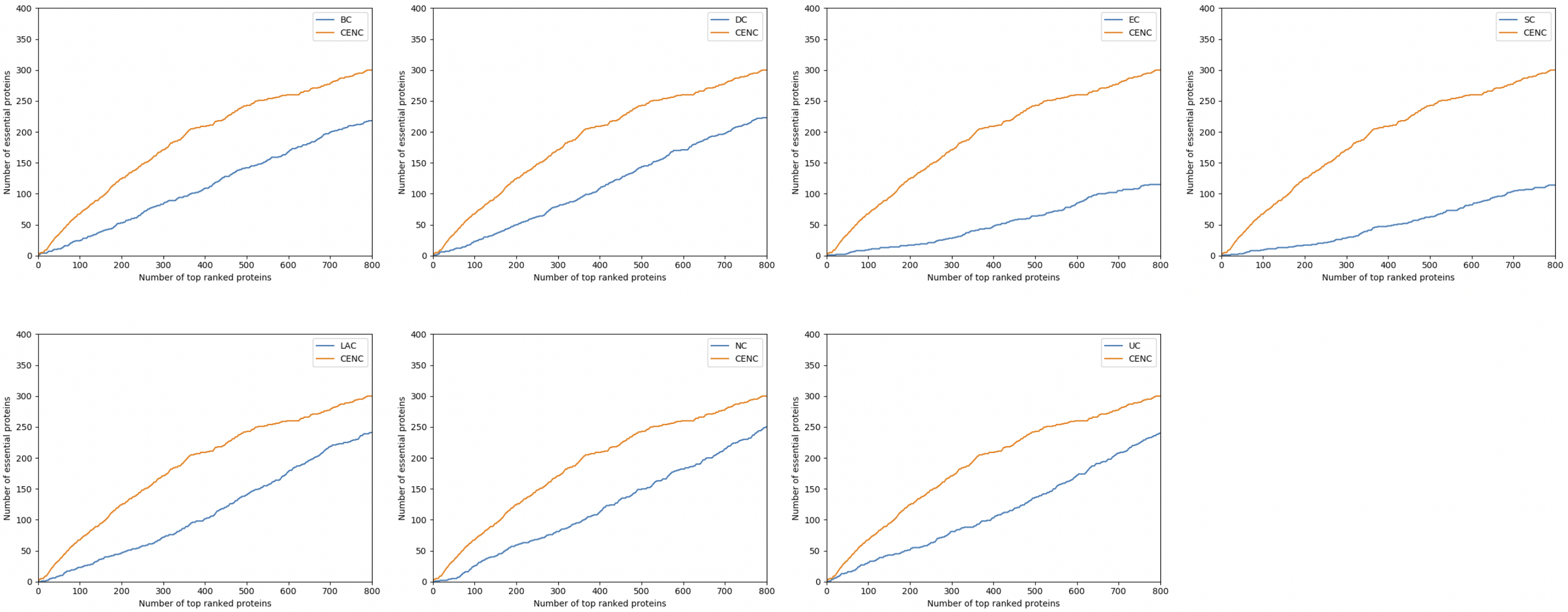}
\caption{The performances of $CENC$ and other seven centrality measures on the MIPS network are evaluated by a jackknife methodology.}
\centering
\end{figure}


\section{Conclusion}\label{Se5}
Essential proteins are crucial for the survival and normal functioning of all organisms. Improving the recognition accuracy of essential proteins is a challenging task. Plenty of scholars devoted themselves to identify essential proteins in terms of the topological features for the whole network, ignoring the importance of complex and biological information. In this paper, on the basis of the mixed clustering coefficient for complexes and edge topology, a new method $CENC$ is proposed. Then two different datasets of MIPS and DIP are applied. The evaluation methods include ``sorting-screening" method, six statistical method, the precision-recall curves, $ROC$ curve, $AUC$ and jackknife method. Then we compare $CENC$ and other seven proposed methods, containing $DC$, $BC$, $EC$, $SC$, $LAC$, $NC$ and $UC$ by using these evaluation methods. It is found that our proposed method of $CENC$ has the ability to improve the accuracy in predicting essential proteins.
\medskip



\begin{thebibliography}{99}
\footnotesize{
\bibitem{1}
Fraser H B, Hirsh A E, et al.,
\newblock Evolutionary Rate in the Protein Interaction Network,
\newblock Science, 296(5568):750-752, 2002.

\bibitem{2}
Xu B, Guan J, Wang Y, et al.,
\newblock Essential protein detection by random walk on weighted protein-protein interaction networks,
\newblock IEEE/ACM Trans Comput Biol Bioinform, PP(99):1-1, 2017.

\bibitem{3}
Winzeler E A, Shoemaker D D, Astromoff A, Liang H, Anderson K, Andre B, et al.,
\newblock Functional characterization of the s. cerevisiae genome by gene deletion and parallel analysis,
\newblock Science, 285 (5429):901-906, 1999.

\bibitem{4}
Wang Y, Sun H, Du W, Blanzieri E, Viero G, Xu Y, et al.,
\newblock Identification of essential proteins based on ranking edge-weights in protein-protein interaction networks,
\newblock PloS One, 9(9):e108716, 2014.

\bibitem{5}
Roemer T, Jiang B, Davison J, Ketela T, Veillette K, et al.,
\newblock  Large-scale essential gene identification in Candida albicans and applications to antifungal drug discovery,
\newblock Mol Microbiol, 50:167-181, 2003.

\bibitem{6}
Cullen L M, Arndt G M,
\newblock Genome-wide screening for gene function using RNAi in mammalian cells,
\newblock Immunol Cell Biol, 83:217-223, 2005.

\bibitem{43}
Estrada E,
\newblock Virtual identification of essential proteins within the protein interaction network of yeast,
\newblock Proteomics, 6(1):35-40, 2006.

\bibitem{44}
Peng W, Wang J, Wang W, et al.,
\newblock Iteration method for predicting essential proteins based on orthology and protein-protein interaction networks,
\newblock BMC Systems Biology, 6(1),2012.

\bibitem{7}
Giaever G, Chu A M, Ni L, et al.,
\newblock SGD: Functional profiling of the saccharomyces cerevisiae genome,
\newblock Nature, 418(6896):387-391, 2002.

\bibitem{8}
Jeong H M, Mason S P, Albert B, et al.,
\newblock Lethality and centrality in protein networks,
\newblock Nature, 411:41-42, 2001.

\bibitem{9}
Zhao B H, Wang J X, Li M, et al.,
\newblock  Prediction of Essential Proteins Based on Overlapping Essential Modules,
\newblock IEEE Transactions on Nanobioscience, 13(4):415-424, 2014.

\bibitem{10}
Hahn M W, Kern A D,
\newblock Comparative genomics of centrality and essentiality in three eukaryotic protein-interaction networks,
\newblock Molecular Biology and Evolution, 22(4):803-806, 2005.

\bibitem{11}
Freeman L C,
\newblock A set of measures of centrality based on betweenness,
\newblock Sociometry, 40(1):35-41, 1977.

\bibitem{12}
Li M , Li W , Wu F X , et al.,
\newblock Identifying essential proteins based on sub-network partition and prioritization by integrating subcellular localization information,
\newblock  Journal of Theoretical Biology, 2018.

\bibitem{13}
Wuchty S, Stadler P F,
\newblock Centers of complex networks,
\newblock Journal of Theoretical Biology, 223(1):45-53, 2003.

\bibitem{47}
Batada N N, Hurst L D, Tyers M,
\newblock Evolutionary and Physiological Importance of Hub Proteins,
\newblock PLoS Computational Biology, 2(7):e88, 2006.

\bibitem{14}
Lin C C, Juan H F, Hsiang J T, Hwang Y C, Mori H, Huang H C,
\newblock Essential core of protein-protein interaction network in escherichia coli,
\newblock Journal of Proteome Research, 8(4):1925-1931, 2009.

\bibitem{15}
Liang H, Li W H,
\newblock Gene essentiality, gene duplicability and protein connectivity in human and mouse,
\newblock Trends in Genetics, 23(8):375-378, 2007.

\bibitem{16}
Estrada E, Juan A,
\newblock Subgraph centrality in complex networks,
\newblock Physical Review E, 71(5):1-9, 2005.

\bibitem{17}
Li M, Wang J, Chen X, et al.,
\newblock A local average connectivity-based method for identifying essential proteins from the network level,
\newblock Computational Biology and Chemistry, 35(3):143-150, 2011.

\bibitem{19}
Bonacich P,
\newblock Power and centrality: a family of measures,
\newblock  American Journal of Sociology, 92(5):1170-1182, 1987.

\bibitem{48}
Nie T , Guo Z , Zhao K , et al.,
\newblock Using mapping entropy to identify node centrality in complex networks
\newblock Physica A, Statistical Mechanics and its Applications, 453:290-297, 2016.

\bibitem{49}
Radicchi F, Castellano C, Cecconi F, et al.,
\newblock Defining and identifying communities in networks,
\newblock Proceedings of the National Academy of Sciences of the United States of America, 101(9):2658-2663, 2003.

\bibitem{45}
Jiang Y , Wang Y , Pang W , et al.,
\newblock CytoNCA: Essential Protein Identification Based on Essential Protein-Protein Interaction Prediction by Integrated Edge Weights,
\newblock The IEEE International Conference on Bioinformatics and Biomedicine. IEEE, 2014.

\bibitem{46}
Wang J , Li M , Wang H , et al.,
\newblock Identification of Essential Proteins Based on Edge Clustering Coefficient,
\newblock IEEE/ACM Transactions on Computational Biology and Bioinformatics, 9(4):1070---1080, 2012.

\bibitem{42}
Tang Y , Li M , Wang J , et al.,
\newblock CytoNCA: A cytoscape plugin for centrality analysis and evaluation of protein interaction networks,
\newblock Biosystems, 127:67-72, 2015.

\bibitem{20}
Stephenson K, Zelen M,
\newblock Rethinking centrality: methods and examples,
\newblock Soc Networks, 11:1-37, 1989.

\bibitem{39}
Li G , Li M , Wang J , et al.,
\newblock United neighborhood closeness centrality and orthology for predicting essential proteins,
\newblock  IEEE/ACM Transactions on Computational Biology and Bioinformatics, 2018.

\bibitem{21}
Zhang Z P, Ruan J S, Gao J Z, et al.,
\newblock Predicting essential proteins from protein-protein interactions using order statistics,
\newblock  Journal of Theoretical Bioligy, 480:274-283, 2019.

\bibitem{22}
Hart G T, Lee I, Marcotte E M.
\newblock A high-accuracy consensus map of yeast protein complexes reveals modular nature of gene essentiality.
\newblock Bmc Bioinformatics, 8(1):236-0, 2007.

\bibitem{23}
Li M, Lu Y, Niu Z, et al.,
\newblock United complex centrality for identification of essential proteins from PPI networks,
\newblock IEEE/ACM Transactions on Computational Biology and Bioinformatics, 14(2):370-380, 2017.

\bibitem{24}
Li M, Zhang H H, Fei Y P,
\newblock Essential protein discovery method based on integration of PPI and gene expression data,
\newblock Journal of Central South University, 44(3):1024-1029, 2013.


\bibitem{25}
Lei X , Zhao J , et al.,
\newblock Predicting essential proteins based on RNA-Seq, subcellular localization and GO annotation datasets,
\newblock  Knowledge-Based Systems, 2018.

\bibitem{26}
Xenarios I, Lukasz S, et al.,
\newblock DIP, the database of interacting proteins: a research tool for studying cellular networks of protein interactions,
\newblock Nucleic Acids Research, 30(1):303-305, 2002.

\bibitem{27}
Zhang R, Lin Y,
\newblock DEG 5.0, a database of essential genes in both prokaryotes and eukaryotes,
\newblock Nucleic Acids Res, 37(suppl 1):D455-D458, 2009.

\bibitem{28}
Wang J, Li M, Wang H, Pan Y,
\newblock  Identification of essential proteins based on edge clustering coefficient,
\newblock Transactions on Computational Biology and Bioinformatics, 9(4):1070-1080, 2012.

\bibitem{29}
Friedel C C, Krumsiek J, Zimmer R,
\newblock International Conference on Research in Computational Molecular Biology,
\newblock Springer-Verlag, 2008.

\bibitem{30}
Pu S, Wong J, Turner B, Cho E, Wodak S J,
\newblock Up-to-date catalogues of yeast protein complexes,
\newblock Nucleic Acids Research, 37(3):825-831, 2009.

\bibitem{31}
Pu S, Vlasblom J, Emili A, et al.,
\newblock  Identifying functional modules in the physical interactome of saccharomyces cerevisiae,
\newblock Proteomics, 7(6):944-960, 2010.

\bibitem{32}
Holman A G, Davis P J, Foster J M, et al.,
\newblock Computational prediction of essential genes in an unculturable endosymbiotic bacterium, wolbachia of brugia malayi,
\newblock Bmc Microbiology, 9(1):1-14, 2009.

\bibitem{33}
Cherry J M, Adler C, Ball C A, et al.,
\newblock SGD: saccharomyces genome database,
\newblock Nucleic Acids Research, 26(1):73-79, 1998.

\bibitem{34}
Li M , Lu Y , Wang J , et al.,
\newblock A Topology Potential-Based Method for Identifying Essential Proteins from PPI Networks,
\newblock IEEE/ACM Transactions on Computational Biology and Bioinformatics, 12(2):372-383, 2015.

\bibitem{35}
Radicchi F , Castellano C , Cecconi F, et al.,
\newblock Defining and identifying communities in networks,
\newblock  Proceedings of the National Academy of Sciences of the United States of America, 101(9):2658-2663, 2003.

\bibitem{36}
Zhu Y, Wu C,
\newblock Identification of essential proteins using improved node and edge clustering coefficient,
\newblock  Proceedings of the 37th Chinese Control Conference, 2018.

\bibitem{37}
Luo J W, Qi Y,
\newblock Identification of essential proteins based on a new combination of local interaction density and protein complexes,
\newblock PLOS ONE, 10(6):e0131418, 2015.

\bibitem{38}
Joy M P, Brock A, Ingber D E, et al.,
\newblock High-betweenness proteins in the yeast protein interaction network,
\newblock Journal of Biomedicine and Biotechnology, 2005(2):96, 2014.

\bibitem{40}
Mewes H W, Amid C, Arnold R, et al.,
\newblock MIPS: analysis and annotation of proteins from whole genomes,
\newblock Nucleic Acids Research, 34(Database issue):169-72, 2004.

\bibitem{41}
Pereira-Leal J B, Benjamin A , Peregrin-Alvarez J M, et al.,
\newblock An Exponential Core in the Heart of the Yeast Protein Interaction Network,
\newblock  Molecular Biology and Evolution, 2015.

\bibitem{50}
P Bradley,
\newblock The Use of the Area Under the ROC Curve in the Evaluation of Machine Learning Algorithms,
\newblock Pattern Recognition, 30:1145-1159, 1996.









}
\end{thebibliography}
\end{document}